\documentclass{emulateapj}
\usepackage{apjfonts}
\usepackage{epsfig,amssymb,amsmath,natbib}
\usepackage{graphicx}
\usepackage{epstopdf}

\begin{document}

\title{A Deeper Look at Faint H$\alpha$ Emission in Nearby Dwarf Galaxies}

\author{
Janice C. Lee\altaffilmark{1,2}, Sylvain Veilleux\altaffilmark{3}, Michael McDonald\altaffilmark{4}, Bryan Hilbert\altaffilmark{1}
}
\altaffiltext{1}{Space Telescope Science Institute, Baltimore, MD; jlee@stsci.edu}
\altaffiltext{2}{Visiting Astronomer, SSC/IPAC, Caltech, Pasadena, CA}
\altaffiltext{3}{Department of Astronomy and Joint Space-Science Institute, University of Maryland, College Park, MD}
\altaffiltext{4}{MIT Kavli Institute for Astrophysics and Space Research, Cambridge, MA}

\begin{abstract}
%203 words; max is 250
%We present deep H$\alpha$ imaging of three nearby dwarf galaxies, carefully selected to optimize observations with the Maryland-Magellan Tunable Filter (MMTF) on the Magellan 6.5m telescope.  An effective bandpass of $\sim$13\AA\ is used, and the images reach 3$\sigma$ flux limits of $\sim$8$\times10^{-18}$ ergs s$^{-1}$ cm$^{-2}$, which is about an order of magnitude lower than standard narrowband observations obtained by the most recent generation of local H$\alpha$ galaxy surveys. This work was motivated by the finding that the H$\alpha$/FUV flux ratio of galaxies systematically declines as global galactic properties such as the star formation rate and stellar mass decrease. 
%Follow-up studies of many of the potential causes for the trends in the H$\alpha$/FUV flux ratio have been performed, but the possibility that previous observations have missed a non-negligible fraction of faint ionized emission in dwarf galaxies has not been investigated.  The MMTF observations reveal both diffuse and structured H$\alpha$ emission (filaments, shells, single-star HII regions) spanning extents up to 2.5 times larger relative to previous observations.  An additional $\sim5-20$\% of H$\alpha$ flux is captured, which does not entirely explain the trends in the H$\alpha$/FUV ratio.  However, the impacts of the newly detected extended flux on our understanding of star formation, the properties of HII regions, and the propagation of ionizing photons warrant further investigation.

We present deep H$\alpha$ imaging of three nearby dwarf galaxies, carefully selected to optimize observations with the Maryland-Magellan Tunable Filter (MMTF) on the Magellan 6.5m telescope.  An effective bandpass of $\sim$13\AA\ is used, and the images reach 3$\sigma$ flux limits of $\sim$8$\times10^{-18}$ ergs s$^{-1}$ cm$^{-2}$, which is about an order of magnitude lower than standard narrowband observations obtained by the most recent generation of local H$\alpha$ galaxy surveys. The observations were originally motivated by the finding that the H$\alpha$/FUV flux ratio of galaxies systematically declines as global galactic properties such as the star formation rate and stellar mass decrease. The three dwarf galaxies selected for study have star formation rates, that when calculated from their H$\alpha$ luminosities using standard conversion recipes, are $\sim$50\% of those based on the FUV.  Follow-up studies of many of the potential causes for the trends in the H$\alpha$/FUV flux ratio have been performed, but the possibility that previous observations have missed a non-negligible fraction of faint ionized emission in dwarf galaxies has not been investigated.  The MMTF observations reveal both diffuse and structured H$\alpha$ emission (filaments, shells, possible single-star HII regions) spanning extents up to 2.5 times larger relative to previous observations.  However, only up to an additional $\sim5$\% of H$\alpha$ flux is captured, which does not account for the trends in the H$\alpha$/FUV ratio.  Beyond investigation of the H$\alpha$/FUV ratio, the impact of the newly detected extended flux on our understanding of star formation, the properties of HII regions, and the propagation of ionizing photons warrant further investigation.

%%The abstract should mention that the percentage of missing flux is larger at higher inclinations.
%%In the abstract, the authors should also state explicitly how much
%%flux was expected to be missing given Halpha/FUV at the SFR for these
%%galaxies. What fraction of that was recovered? The sentence in the
%%abstract "which does not entirely explain the trends" is misleading:
%%in the discussion, the authors show that the recovered flux has little
%%impact.

\end{abstract}


\section{Introduction} 

Observations of the diffuse (warm) ionized gas in galaxies have been essential for advancing our understanding of star formation feedback processes, and the structure and phase balance of the ISM.  The DIG (or warm ionized medium -- WIM) is commonly studied with observations of the H$\alpha$ emission-line.  From the first detection of the Galactic ``Reynolds Layer" (Reynolds 1971; Reynolds et al. 1984), to the discovery of DIG in the haloes of edge-on galaxies and throughout galaxy disks, intricate geometries of loops, shells, filaments and chimneys -- the fossil remnants of past deposition of radiative and mechanical energy into the ISM by massive stars and stellar death -- have been found  (e.g., Dettmar 1990; Rand et al., 1990; Monnett 1971; Hunter \& Gallagher 1990; Ferguson et al. 1998; Haffner et al. 2009 and references therein).  The contribution of the DIG, whether in such structures or in an unresolved or genuinely diffuse component, to the total H$\alpha$ luminosity of galaxies is substantial.  On average, about half of the integrated H$\alpha$ emission is measured outside discrete HII regions (e.g., van Zee 2000; Oey et al. 2007, and references therein).

Here, we present the results of deep H$\alpha$ imaging observations designed to significantly surpass the flux limits that characterize the most recent generation of local H$\alpha$ galaxy surveys (e.g., Meurer et al. 2006; Kennicutt et al. 2008, Gavazzi et al. 2012; Boselli et al. 2015), and past studies which have played a major role in shaping our view of the DIG in nearby galaxies.  Our objective was to perform a simple detection experiment for faint H$\alpha$ in the outskirts of dwarf galaxies.  The observations were originally motivated by the finding that the H$\alpha$/FUV flux ratio of galaxies systematically declines as global galactic properties such as the star formation rate (SFR) and stellar mass decrease (e.g., Boselli et al. 2009; Meurer et al. 2009; Hunter et al. 2010; Lee et al 2009; 2011a, b and references therein).  The behavior of the H$\alpha$/FUV ratio was studied over five orders of magnitudes in SFR to $\sim$10$^{-4}$ M$_{\odot}$ yr$^{-1}$ by Lee et al. (2009).  Above SFR$\sim$0.1 M$_{\odot}$ yr$^{-1}$, the average ratio is constant, but was found to be lower by a factor of two at $\sim$10$^{-3}$ M$_{\odot}$ yr$^{-1}$, and by an order of magnitude at $\sim$10$^{-4}$ M$_{\odot}$ yr$^{-1}$.  Possible causes include bursty or otherwise non-constant star formation histories, variations in the stellar initial mass function (IMF) for massive stars, stochastic sampling of the massive star IMF, undetected H$\alpha$ diffuse emission, and the escape of ionizing photons into the intergalactic medium.  The behavior of the ratio is likely due to a combination of effects, and subsequent work has investigated the possibilities in further detail to attempt to explain the behavior of the H$\alpha$/UV ratio in dwarf galaxies as well as of the low surface brightness outer disks of spiral galaxies (e.g., Fumagalli et al. 2011; Weisz et al. 2012; Eldridge et al. 2012; Relano et al. 2012; Weidner et al. 2013; da Silva et al. 2014).  Here, we report on our effort to examine the topic that has not been as well-studied as the others: that the low gas density environments typical of dwarf galaxies might allow ionizing photons to travel far from their parent HII regions, and result in diffuse H$\alpha$ emission that falls beneath the detection thresholds of typical narrowband observations of nearby galaxies.

To perform this experiment, we have used the Maryland-Magellan Tunable Filter (MMTF; Veilleux et al. 2010) with the Inamori Magellan Areal Camera and Spectrograph (IMACS; Dressler et al. 2006), and improve upon major observational parameters which affect the flux sensitivity by factors of at least a few.  The H$\alpha$ galaxy surveys on which the UV/H$\alpha$ studies have been based have typically used $\sim$2 meter class telescopes, R$\sim$100 narrowband (NB) filters ($\sim$70\AA), and NB exposure times of $\sim$20 minutes.  With our MMTF observations, we gain the 6.5m aperture of the Magellan-Baade telescope, employ an effective bandpass of $\sim$13\AA, and lengthen the narrowband integration to $\sim$3 hours.  The resulting H$\alpha$ images reach flux limits which are 8-9 lower than previous imaging.  Further, this work also addresses potentially large uncertainties in continuum subtraction that can particularly affect dwarf galaxies.  To estimate the stellar continuum flux density, the standard approach is to scale a continuum image (which is often taken in a broadband $R$ filter for efficiency) to match the average continuum level of the galaxy in the narrowband image.   The scaling is calibrated using pure continuum sources (e.g., foreground stars, and areas within the target galaxy that are thought to be continuum dominated). However, uncertainties (20-60\%; Oey et al. 2007; Kennicutt et al. 2008) arise from differences in the continuum slope between the target galaxies (dwarf galaxies are among the bluest known locally) and the calibrating sources, and may lead to over-subtraction of the DIG.  Our work with the MMTF essentially removes these uncertainties since the continuum image is taken at a wavelength close to H$\alpha$ with the same narrow bandpass used for the on-line image; moreover, the contribution of the continuum is small due to the narrowness of the bandpass.  Further details on the observations and sample selection are given in \S\ 2, while the data reduction is summarized in \S\ 3.  The results of the MMTF observations are presented in \S\ 4, and summarized in \S\ 5.

%Finally, with Magellan we also gain the superb sub-arcsecond seeing at Las Campanas, which allows possible additional structure in the DIG to be resolved.

\section{Observations}

The nearby dwarf galaxies CGCG035-007, IC559, and UGC5456 were selected for observation from the 11 Mpc H$\alpha$ UV Galaxy and Spitzer Local Volume Surveys (11HUGS/LVL; Kennicutt et al. 2008; Lee et al. 2009; Dale et al. 2009; Lee et al. 2011; Cook et al. 2014).  The galaxies were chosen to:
\begin{enumerate}
\item exhibit low H$\alpha$-to-FUV ratios ($\sim$0.5 when H$\alpha$ and FUV are converted to extinction-corrected SFRs using the relationships in Kennicutt 1998; see Lee et al. 2009),
\item have blue UV colors (FUV-NUV$<$0.4) to ensure that the low H$\alpha$-to-FUV ratios are not primarily due to excess UV emission from old stellar populations (i.e., UV ``upturn''),
\item have small angular extents ($\sim$2\arcmin) to fit within the monochromatic spot on a single IMACS detector for the chosen etalon plate spacing (see Section 3),
\item span a range of disk inclinations, to enable study of faint H$\alpha$ emission both external to and within the disk plane,
\item have a minimum of bright foreground stars within a 7\arcmin\ radius to minimize potential ghosting effects (section 4.3.2, Veilleux et al. 2010), and
\item be observable in the southern sky in the first half of the year.
\end{enumerate}
Salient properties of the galaxies are summarized in Table 1.  Figure 1 shows the locations of the three galaxies on a plot comparing H$\alpha$ and FUV luminosities (extinction corrected) of the 11HUGS/LVL sample as in Lee et al. (2009).

Observations were conducted between 2010 March 11-15.  New moon occurred on the night of March 15.  Observing conditions were photometric through the run.  The seeing was generally quite good (0\farcs6-0\farcs7), though there were increases up to $\sim$1\farcs2 on the later halves of the first two nights. 

The MMTF uses a low-order Fabry-Perot etalon and is mounted in the IMACS f/2 camera, which has an 8K $\times$ 8K mosaic CCD with a pixel scale of 0\farcs2.  The observations are taken in "staring mode," where the etalon gap is fixed, resulting in a single narrowband image at each point in the field.  However, the central wavelength of the resultant narrowband slowly varies with distance from the optical axis.  The selected etalon plate spacing ($Z_{coarse}=+1$) yields a 4.8\arcmin diameter field-of-view over which the central wavelength of the bandpass does not change by more than 3 \AA.  The bandpass FWHM is 8.1 \AA, and the profile shape is well characterized by a Lorentzian function, which has an effective bandpass (the integral of the profile divided by its peak) of $\Pi/2 \times 8.1$ \AA\ (12.7 \AA).

The etalon was tuned to capture redshifted H$\alpha$ emission at the recessional velocities of the galaxies.  Continuum images, both to the blue and to the red of H$\alpha$, were obtained with the MMTF using the same etalon parameters as the H$\alpha$ observations, as well as with medium-band filters that do not contain strong emission lines (MMTF6399-206 and MMTF7045-228).  The drift of the central wavelength of the MMTF bandpass was minimized by taking the observations at a constant gravity angle, and by re-calibrating the etalon approximately every hour.  As a result, the field gradually rotates on the detector plane over the span of the night.  The etalon was tuned immediately prior to most H$\alpha$ exposures, with the result that the central wavelength was held constant to within $\pm$0.3\AA.  Continuum exposures followed most H$\alpha$ exposures, but were taken without re-calibrating, so the drift of the central wavelength is somewhat higher, but still within $\pm$1\AA.  Observations of spectrophotometric standard stars from  Massey et al. (1988) were taken at the beginning and end of each night.

As summarized in Table 2, total exposure times per object were $\sim$ 3 hours through the MMTF H$\alpha$ bandpass, $\sim$40 minutes through each of MMTF continuum bandpasses, and at least 10 minutes in each of the medium-band filters.

\section{Reduction}
Bias subtraction, flat fielding, and sky subtraction were performed using version 1.4 of the MMTF data reduction pipeline\footnote{available at http://www.astro.umd.edu/~veilleux/mmtf/datared.html}, with modifications to mitigate source contamination of the overscan region, and to improve the sky determination via source masking.

Subtraction of the stellar continuum from the H$\alpha$ images was performed as follows.  To minimize the impact of the subtraction on the overall depth of the imaging, medium-band images which bracket the MMTF H$\alpha$ observations were used.  The factors needed to scale the medium-band images to the level of the continuum in the H$\alpha$ observations were found from the flux ratios of unsaturated, well-exposed stars in common between each pair of MMTF continuum and medium-band images.  As a check, the ratios of the responses (computed from the standard star observations) for the medium-band and MMTF H$\alpha$ images were also calculated.  The values agree to 5\%.  The scaled medium-band images were then combined and subtracted from the stacked H$\alpha$ image.  This process however, does not correct for underlying stellar H$\alpha$ absorption, which is typically $\sim$2\AA\ in equivalent width.  Thus, the integrated fluxes reported at the bottom of Table 1 are formally lower limits, but iterative subtraction of progressively fainter continuum images show that the H$\alpha$ fluxes do not increase by more than a few percent before the images appear to be clearly under-subtracted.

\section{Results}

The results of the MMTF observations are first compared with previous H$\alpha$ imaging in Figures 1, 2, and 3.  A 2$\times$3 panel figure is shown for each dwarf galaxy.  In each figure, the first two rows show the same pair of H$\alpha$ images: one from the 11HUGS/LVL survey on the left, and the deeper MMTF image on the right.  The top pair of images are shown on a logarithmic scale and are stretched to highlight bright HII regions in the inner disk, while the middle pair of images are shown on a linear scale and are stretched to emphasize faint emission and extended DIG.  Contours are chosen to complement the stretch of the images, so that structures over a full range of surface brightnesses can be examined in each individual panel.  Contours at 1 and 3$\sigma$ of the background outline the full extent of detectable emission in the top pair of images, while 10, 30, and 100$\sigma$ contours are drawn on the inner regions in the middle pair of images.  Whereas the 1$\sigma$ contours correspond to 3-4 $\times 10^{-17}$ ergs s$^{-1}$ cm$^{-2}$ arcsec$^{-2}$ in the 11HUGS/LVL images, they reach 8-9 times deeper in the MMTF observations, to 4-5 $\times 10^{-18}$ ergs s$^{-1}$ cm$^{-2}$ arcsec$^{-2}$.  Thus, the shape of the 1$\sigma$ 11HUGS/LVL contours is similar to the 10$\sigma$ MMTF contours for each galaxy. The bottom two panels in each figure show the MMTF contours again, but overlaid on the medium-band image used for continuum subtraction to illustrate the morphology of the H$\alpha$ emission relative to the stellar disk.

It is immediately clear that the MMTF observations reveal H$\alpha$ emission that is far more extended and finely structured than previously seen in the 11HUGS/LVL data.   The nature of the faint emission detected with the MMTF depends on the inclination of the disk -- on whether H$\alpha$ is observed external to the disk plane or in the outer periphery of the disk.  We first describe notable morphological features of the emission, and then discuss the measurement and comparison of the fluxes from the 11HUGS/LVL and MMTF images.

\subsection{Morphology of the H$\alpha$ Emission}

UGC5456 (Figure 1) is highly inclined ($i=80\degr$; Makarov et al. 2014), and DIG is detected $\sim$2.5 times further beyond the disk mid-plane relative to the 11HUGS/LVL image.  The DIG is more extended to the northeast, rising to 0\farcm9 above the midplane, and to 0\farcm6 to the southwest.  These angular distances correspond to 1 kpc and 0.7 kpc respectively for the 11HUGS/LVL adopted distance of 3.8 Mpc (Kennicutt 2008).  If a more recent TRGB distance of 10.5 Mpc from the Extragalactic Distance Database\footnote{\url{http://edd.ifa.hawaii.edu/}} is used, the physical extents would scale accordingly by a factor of 2.8.  The emission in the northeast may extend to the edge of the high signal-to-noise region of the MMTF image, which coincides with the top of the image (Figure 2, middle right); the potential emission is not captured by the 1$\sigma$ contours due to the 4 pixel smoothing, but clearly may only be noise.  The higher resolution of the MMTF image (PSF FWHM of 0\farcs7 relative to 1\farcs2; corresponding to 13 and 22 pc at 3.8 Mpc) shows that the faint H$\alpha$ emission is highly structured, with well-defined loops and filaments surrounding the bright HII regions (Figure 1, top right).  Similar but more diffuse structures are also evident in the galaxy halo, and numerous chimney-like spokes, possibly highlighting low column density pathways for Lyman continuum photon propagation into the halo (e.g., Zastrow et al. 2011, 2013), can be identified just beyond the 10$\sigma$ contour (Figure 2, middle right).  At the southern tip of the galaxy, an additional low surface brightness HII region is detected (Figure 2, middle right), and a single irregular region in the 11HUGS/LVL image is clearly resolved into 7 objects.  The observed (i.e., no extinction corrections applied) H$\alpha$ luminosity of these regions range from 5$\times$10$^{35}$ ergs s$^{-1}$ to 2$\times$10$^{36}$ ergs s$^{-1}$, which corresponds to the ionizing fluxes of single B0V-O9V stars (Smith et al. 2002), assuming Case B recombination and that the HII regions are radiation bounded.  If the galaxy is instead at 10.5 Mpc, the H$\alpha$ luminosities would instead map to more massive stars in the O8V-O7V range.
  
In contrast, IC559 (Figure 2) is the least inclined of the three galaxies in our sample ($i=35\degr$), and although the MMTF image again reveals loops and filaments in the immediate vicinity of the bright HII regions (Figure 2, top), the extraplanar DIG found in UGC5456 is not seen.  Rather, the newly detected H$\alpha$, which is twice as extended relative to previous observations, takes the form of many faint HII regions which surround the periphery of the 10$\sigma$ (1$\sigma$) contour of the MMTF (11HUGS/LVL) image (Figure 2, middle).   The regions are even less luminous than those in UGC5456 (assuming a distance of 4.9 Mpc; Kennicutt 2008). Their H$\alpha$ luminosities range from 8$\times$10$^{34}$ ergs s$^{-1}$ to 5$\times$10$^{35}$ ergs s$^{-1}$; the low end of this range corresponds to the ionizing flux of a single B0.5V star (Smith et al. 2002).

Finally, CGCG035-007 (Figure 3) is at an inclination intermediate between the two previous galaxies ($i=50\degr$).  The new faint emission and structures detected in the MMTF image are similar in nature to that in IC559, although the loops and filaments surrounding the HII regions are not as prominent.  Within the 3$\sigma$ contours of the 11HUGS/LVL image, at least 20 well-defined point-like regions can now be identified (Figure 3, top), and have H$\alpha$ luminosities in the range of the those found at the edges of the H$\alpha$ emission in IC559.  Additional faint HII regions can also seen outside the 10$\sigma$ contour of the MMTF image (Figure 3, middle) mostly on the western side of the galaxy.

\subsection{Comparison of Fluxes}

Despite the much larger extent of the faint H$\alpha$ emission uncovered in the MMTF observations, its contribution to the integrated flux is small, as might be expected.  It is straightforward to compute the fractional contribution using relative measurements in two elliptical apertures: a smaller aperture used to measure the total flux in the 11HUGS/LVL images, and a larger one that also covers the more extended emission found in the MMTF data.  These apertures are shown in the top panels in each figure, and the corresponding measurements, along with the percent difference are given in Table 1.  Measurements of  11HUGS/LVL H$\alpha$ images with the larger apertures do show 1-6\% greater flux compared to those with the smaller apertures, but this difference is insignificant given the 1$\sigma$ uncertainties (12-18\%).  Analogous measurements on the MMTF images yield 3-7\% greater flux detected at the 1$\sigma$ level in the outer regions (uncertainties are 4\%).  

The fluxes measured with the same aperture on the 11HUGS/LVL and MMTF datasets can also be cross-compared.  However, such a comparison is less straightforward since the 11HUGS/LVL imaging includes flux from the [NII]$\lambda\lambda$6548,84 lines, and is subject to uncertainties in its removal.  The comparison would also be affected by potential systematic offsets in the zeropoint scales, which is not captured in the reported errors.  

Regarding the [NII] lines, in Kennicutt et al. (2008) the [NII]$\lambda\lambda$6548,84 emission was estimated to be 3-5\% of the H$\alpha$ flux based on an empirical scaling relationship with the blue absolute magnitude of the galaxy.  However, spectroscopic measurements taken through a 3\arcsec\ fiber are available from SDSS DR12.  The SDSS spectra yield slightly higher [NII]$\lambda\lambda$6548,84/H$\alpha$ ratios: 0.08, 0.07, and 0.12 for UGC5456, IC559 and CGCG035-007.   In Table 1, both the observed H$\alpha$+[NII] flux from 11HUGS/LVL and an [NII]-corrected flux (based on the SDSS spectroscopy) are given.  

Comparison of the 11HUGS/LVL and MMTF H$\alpha$ photometry measured in the smaller (larger) apertures for each of the galaxies shows that the fluxes are within 2-7\% (5-7\%) of each other, and within the uncertainties.  However, the MMTF fluxes of all three galaxies are all higher.  Thus, systematic offsets due to the calibration of the zeropoint and continuum subtraction of the 11HUGS/LVL data are limited to this level.

\section{Discussion and Conclusions}

We have obtained deep MMTF H$\alpha$ observations of three nearby dwarf galaxies with a range of inclination angles to examine the possibility that previous standard narrowband observations may have missed a component of the integrated H$\alpha$ flux hidden in extended low surface brightness DIG.  The MMTF observations reach 3$\sigma$ flux limits of $\sim$8$\times10^{-18}$ ergs s$^{-1}$ cm$^{-2}$, which is about an order of magnitude lower than standard narrowband observations obtained by the most recent generation of local H$\alpha$ galaxy surveys.  If the three galaxies studied here are representative of the broader nearby dwarf galaxy population, then the published integrated H$\alpha$ fluxes from such surveys should not underestimate the total flux by more than $\sim$5\%.  

The initial motivation for this work was to investigate a potential cause of the systematically low values of the H$\alpha$/UV flux in dwarf galaxies.  The value of SFR(H$\alpha$)/SFR(FUV) for CGCG035-007, IC559, and UGC5456 are 0.53, 0.52, and 0.74\footnote{Values in Lee et al. (2009, 2011) have been updated for both the SDSS DR12 spectroscopic [NII]/H$\alpha$ ratios used here, and a correction for an error in the 11HUGS/LVL photometry for UGC5456 as described in Table 1.}, respectively.  The new measurements based on the MMTF H$\alpha$ imaging do not change these values significantly.  Thus, unaccounted H$\alpha$ flux in a faint extended component surrounding dwarf galaxies is not likely a cause of the trends in the H$\alpha$/UV ratio.  However, there are many galaxies with H$\alpha$/FUV ratios lower than those studied here (Figure 1).  It may be worth checking this conclusion on a few such objects, although stochastic sampling of the stellar initial mass function combined with non-constant star formation may be able to explain most of the trend (Fumagalli et al. 2011; da Silva et al. 2014; Krumholz et al. 2015).

%For CGCG035-007 and IC559 which are at modest inclinations, only an additional 5\% of H$\alpha$ flux is measured beyond an elliptical aperture enclosing the 1$\sigma$ contours of a standard narrowband H$\alpha$+[NII] image taken by 11HUGS/LVL, and clearly will not have a significant impact on the H$\alpha$/UV.  However, in the highly inclined UGC5456,  $\sim$20\% greater flux is measured which increases the ratio to 0.78.  About half of the additional flux is due to extended extra-planar emission.  It is possible that such underestimation of the integrated H$\alpha$ flux could be a systematic issue for edge-on dwarfs, and deeper observations of a larger sample of such galaxies would be interesting in this regard.
%Overall, however, the additional extended flux detected in the MMTF dwarf galaxy observations is not sufficient to explain the trends in the H$\alpha$/UV ratio.  

Beyond examination of the H$\alpha$/UV ratio, the MMTF observations have revealed a wealth of structure in newly detected diffuse emission many scale heights above the disk plane (UGC 5456; Figure 1), as well as numerous faint, possibly single-star HII regions within the plane that warrant further investigation (IC559, CGCG035-007; Figure 2 and 3). Study of the properties of the faintest HII regions, coupled with parsec-resolution multi-wavelength UV-optical imaging from HST (e.g., Legacy ExtraGalactic Ultraviolet Survey, LEGUS; Calzetti et al. 2014) will provide insight into the ages and masses of the ionizing sources.  In particular, HST imaging is available for IC559 from LEGUS, which show single point sources within the faint HII regions in the outskirts of the galaxy.  Such a study will provide insight into the questions of whether massive stars can form in relative isolation, without an accompanying cadre of lower mass stars, and whether the process of star formation may be different outside stellar clusters in the ``field" (e.g., Tremonti et al. 2001; Whitmore et al. 2011; Oey et al. 2013).  Further, comparison of the MMTF H$\alpha$ images with neutral HI gas maps (e.g., FIGGS; Begum et al. 2008) will enable study of the star formation law and Lyman continuum photon ISM propagation at the low density frontier.  For example, close inspection of the MMTF observations of UGC5456 indicates that diffuse, but structured H$\alpha$ may extend past the northeastern edge of the high signal-to-noise region of the image.  It is interesting to examine whether such diffuse emission can be found to the edge of the HI disk, as it has been suggested that a non-negligible fraction of Lyman continuum photons may escape dwarf galaxies with low SFRs (e.g., Melena et al. 2009; Hunter et al. 2010). Spectroscopy across this region and beyond will be needed to confirm the extent of the H$\alpha$ emission.  Finally, to better understand the underlying ionization sources, excitation mechanisms and optical depths, measurement of emission-line ratios (e.g., [SII]$\lambda\lambda$6717, 6731/[OIII]$\lambda$5007 to determine the ionization parameter; e.g., Pellegrini et al. 2012; Zastrow et al. 2013) are needed.

\acknowledgements
S.V. and MM acknowledge partial support from NSF grants AST-0606932 and AST-1009583.
We thank the referee for a constructive report that helped to improve this paper.
%%%%%%%%%%%%%%%

{\it Facilities:} \facility{Magellan:Baade (IMACS)}

\newpage

\begin{deluxetable}{cccc}\tablecolumns{4}
\tablecaption{Galaxy Properties \& H$\alpha$ Photometry}
\tablewidth{0pc}
\tablehead{
\colhead{}&
\colhead{CGCG035-007}&
\colhead{IC559}&
\colhead{UGC5456}\\
}
\startdata
$\alpha$, $\delta$\tablenotemark{a}				& 09h 34m 44.7s, 	& 09h 44m 43.9s, 	& 10h 07m 19.6s, \\
								& +06d 25m 32s 	& +09d 36m 54s	& +10d 21m 43s\\
$cz$\tablenotemark{a} \\ 
$\big[$km s$^{-1}$$\big]$		& 574$\pm$38 		&  541$\pm$ 7		& 544$\pm$ 3\\
\\
$D$\tablenotemark{b} \\
$\big[$Mpc$\big]$			& 5.2				& 4.9				& 3.8\\
\\
inclination\tablenotemark{c}               			& 50\degr			& 36\degr			& 80\degr\\
\\
$SFR_{FUV}$\tablenotemark{d}\\
$\big[$log M$_{\odot}$ yr$^{-1}$$\big]$	& 	-2.5				&	-2.4				&-1.9\\
\\

\multicolumn{1}{c}{Smaller Aperture Measurements}\\
$f_{H\alpha+[NII]}$ (11HUGS/LVL) \tablenotemark{e}&&&\\
$\big[$log ergs s$^{-1}$ cm$^{-2}$$\big]$ & -13.23$\pm$0.05 & -13.07$\pm$0.03 & -12.23$\pm$0.05 \\
\\
$f_{H\alpha}$ (11HUGS/LVL) \tablenotemark{f}&&&\\
$\big[$log ergs s$^{-1}$ cm$^{-2}$$\big]$ & -13.27$\pm$0.05 & -13.10$\pm$0.03 & -12.26$\pm$0.05 \\
\\
$f_{H\alpha}$ (MMTF) \tablenotemark{g} &&&\\
$\big[$log ergs s$^{-1}$ cm$^{-2}$$\big]$ & -13.24$\pm$0.01 & -13.08$\pm$0.01 & -12.25$\pm$0.01 \\
\\
\multicolumn{1}{c}{Larger Aperture Measurements}\\
$f_{H\alpha+[NII]}$ (11HUGS/LVL) \tablenotemark{h}&&&\\
$\big[$log ergs s$^{-1}$ cm$^{-2}$$\big]$ & -13.21$\pm$0.06 & -13.05$\pm$0.04 & -12.22$\pm$0.05 \\
\\
$f_{H\alpha}$ (11HUGS/LVL) \tablenotemark{f}&&&\\
$\big[$log ergs s$^{-1}$ cm$^{-2}$$\big]$ & -13.25$\pm$0.06 & -13.08$\pm$0.04 & -12.25$\pm$0.05 \\
\\
$f_{H\alpha}$ (MMTF) \tablenotemark{i} &&&\\
$\big[$log ergs s$^{-1}$ cm$^{-2}$$\big]$ & -13.23$\pm$0.02 & -13.06$\pm$0.02 & -12.22$\pm$0.02 \\
\\
\multicolumn{1}{c}{Percent difference}\\
$f_{H\alpha}$ (11HUGS/LVL)  & 6\%$\pm$18\% & 5\%$\pm$12\% & 1\%$\pm$17\% \\
\\
$f_{H\alpha}$ (MMTF)  & 3\%$\pm$4\% & 4\%$\pm$4\% & 7\%$\pm$4\% \\

\enddata
\tablenotetext{a}{Right acensions, declinations, and recessional velocities taken from the NASA Extragalactic Database (NED). \url{https://ned.ipac.caltech.edu/}}
\tablenotetext{b}{Distances taken from compilation used and documented in Kennicutt et al. (2008).  Distances for CGCG035-007 and IC559 are based on a Local Group flow model, while a TRGB distance is adopted for UG5456.  Uncertainty in the distance for UGC5456 is discussed in Section 4.}
\tablenotetext{c}{Inclinations taken from the Hyperleda database (Makarov et al. 2014). \url{http://leda.univ-lyon1.fr/}}
\tablenotetext{d}{Extinction corrected star formation rates based on integrated FUV fluxes from Lee et al. (2009).}
\tablenotetext{e}{Observed, integrated H$\alpha+$[NII] fluxes from the 11HUGS/LVL survey (Kennicutt (2008). The value reported here for UGC5456 corrects an error in the calculation of the previously published flux (-12.32).  Specifically, the source flag given in Table 3 of Kennicutt et al. 2008 was incorrectly set to 1.3, and resulted in the application of an inappropriate zeropoint.  The source flag should be 1.1. \url{http://irsa.ipac.caltech.edu/data/SPITZER/LVL/}}
\tablenotetext{f}{Contribution of the [NII] doublet flux removed using line ratio measurements from SDSS DR12 as discussed in Section 4.}
\tablenotetext{g}{Observed H$\alpha$ fluxes measured from the MMTF observations in the smaller aperture shown in the top panels of Figures 1-3.}
\tablenotetext{h}{Observed H$\alpha$ fluxes measured from the 11HUGS/LVL observations in the larger aperture shown in the top panels of Figures 1-3.}
\tablenotetext{i}{Observed H$\alpha$ fluxes measured from the MMTF observations within the larger aperture shown in the top panels of Figures 1-3.}
\label{tab:properties}
\end{deluxetable}

\begin{deluxetable}{rccrcccccc}\tablecolumns{10}
\tablewidth{0pc}
\tablecaption{Observations}

\tablehead{
\colhead{(1)}&
\colhead{(2)}&
\multicolumn{2}{c}{(3)} &
\multicolumn{2}{c}{(4)} &
\multicolumn{2}{c}{(5)}&
\colhead{(6)} &
\colhead{(7)} \\
&
&
\multicolumn{2}{c}{MMTF(H$\alpha$)} &
\multicolumn{2}{c}{MMTF(blue cont)} &
\multicolumn{2}{c}{MMTF(red cont)}&
\colhead{MB6399}&
\colhead{MB7045}\\
\colhead{Name}&
\colhead{$6562.80 (1+z)$}&
\multicolumn{2}{c}{8.1\AA} &
\multicolumn{2}{c}{8.1\AA} &
\multicolumn{2}{c}{8.1\AA}&
\colhead{6399\AA/206\AA} &
\colhead{7045\AA/228\AA} \\ 
&
\colhead{} &
\colhead{$\lambda_c$}&
\colhead{$t_{exp}$} &
\colhead{$\lambda_c$}&
\colhead{$t_{exp}$} &
\colhead{$\lambda_c$}&
\colhead{$t_{exp}$} &
\colhead{$t_{exp}$} &
\colhead{$t_{exp}$}\\
\colhead{}&
\colhead{[\AA]}&
\colhead{[\AA]}&
\colhead{[s]}&
\colhead{[\AA]}&
\colhead{[s]}&
\colhead{[\AA]}&
\colhead{[s]}&
\colhead{[s]} &
\colhead{[s]} 
}
\startdata
CGCG035-007	& 6575.6	  & 6575.6 & 9x1200 &6539.7 & 2x1200 &6614.6& 2x1200 & 3x600 & 2x600\\
IC559		& 6574.8	  & 6575.7 & 10x1200 &6544.6& 2x1200 &6614.5& 3x1200 & 2x300 & 3x300\\
UGC5456		& 6574.9	  & 6575.7& 9x1200 &6542.2& 2x1200 &6612.7& 2x1200 & 3x600 & 5x600\\
%UGCA307		& 6581.0	  & 6580.7& 4x1200 &6544.0& 1x1200 &6618.3& 1x1200 & 2x600 & 2x600\\
%UGC8055		& 6576.5	  & 6575.7&  12x1200 &6544.0& 3x1200 &6515.8& 3x1200 & 3x600 & 4x600\\
\enddata
\tablecomments{(1) NED Primary Object Name (2) Wavelength at which H$\alpha$ should be observed given recessional velocity in Table 1. Errors in $cz$ translate into an uncertainty of 0.8 \AA\ for CGCG035-007, and less than 0.2 \AA\ for all other objects. (3) Central wavelength at position of target on focal plane 500 pixels from optical axis, and exposure time for MMTF line observations. FWHM of bandpass is given in the table header.  (4) Central wavelength at position of target on focal plane 500 pixels from optical axis, and exposure time for MMTF continuum observations taken $\sim$40\AA\ to the blue. FWHM of bandpass is given in the table header. (5) Central wavelength at position of target on focal plane 500 pixels from optical axis, and exposure time for MMTF continuum observations taken $\sim$40\AA\ to the red. FWHM of bandpass is given in the table header. (6) Exposure time for continuum observations taken with a conventional medium-band filter to the blue of H$\alpha$.  Central wavelength and FWHM of bandpass are given the table header.  (7) Exposure time for continuum observations taken with a conventional medium-band filter to the red of H$\alpha$.  Central wavelength and FWHM of bandpass os given the table header.}
\label{tab:observations}
\end{deluxetable}

\begin{figure}
\epsscale{0.8}
\plotone{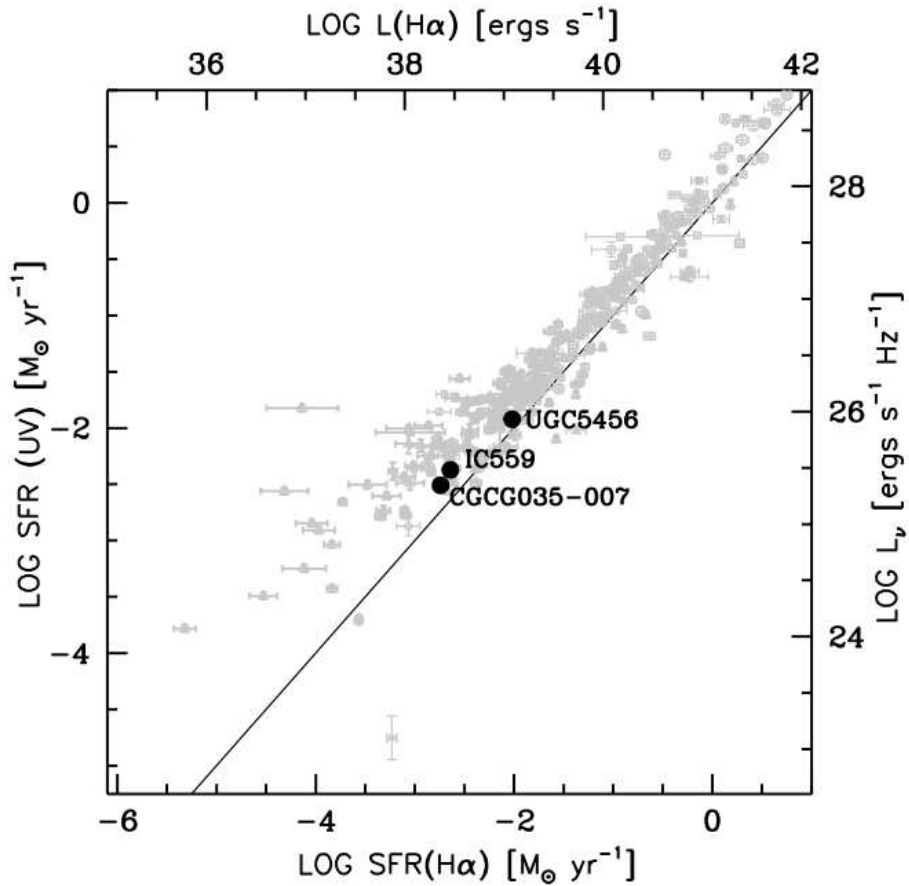}
\caption{Locations of the three galaxies selected for MMTF H$\alpha$ imaging on a plot comparing extinction corrected H$\alpha$ and FUV luminosities of the 11HUGS/LVL sample (gray symbols) as in Lee et al. (2009).  Axes with corresponding star formation rates based on the linear conversion recipes of Kennicutt (1998) are also shown. The solid line shows the one-to-one correspondence between the H$\alpha$ and FUV SFRs.}
\end {figure}

\begin{figure}
\epsscale{0.9}
\plotone{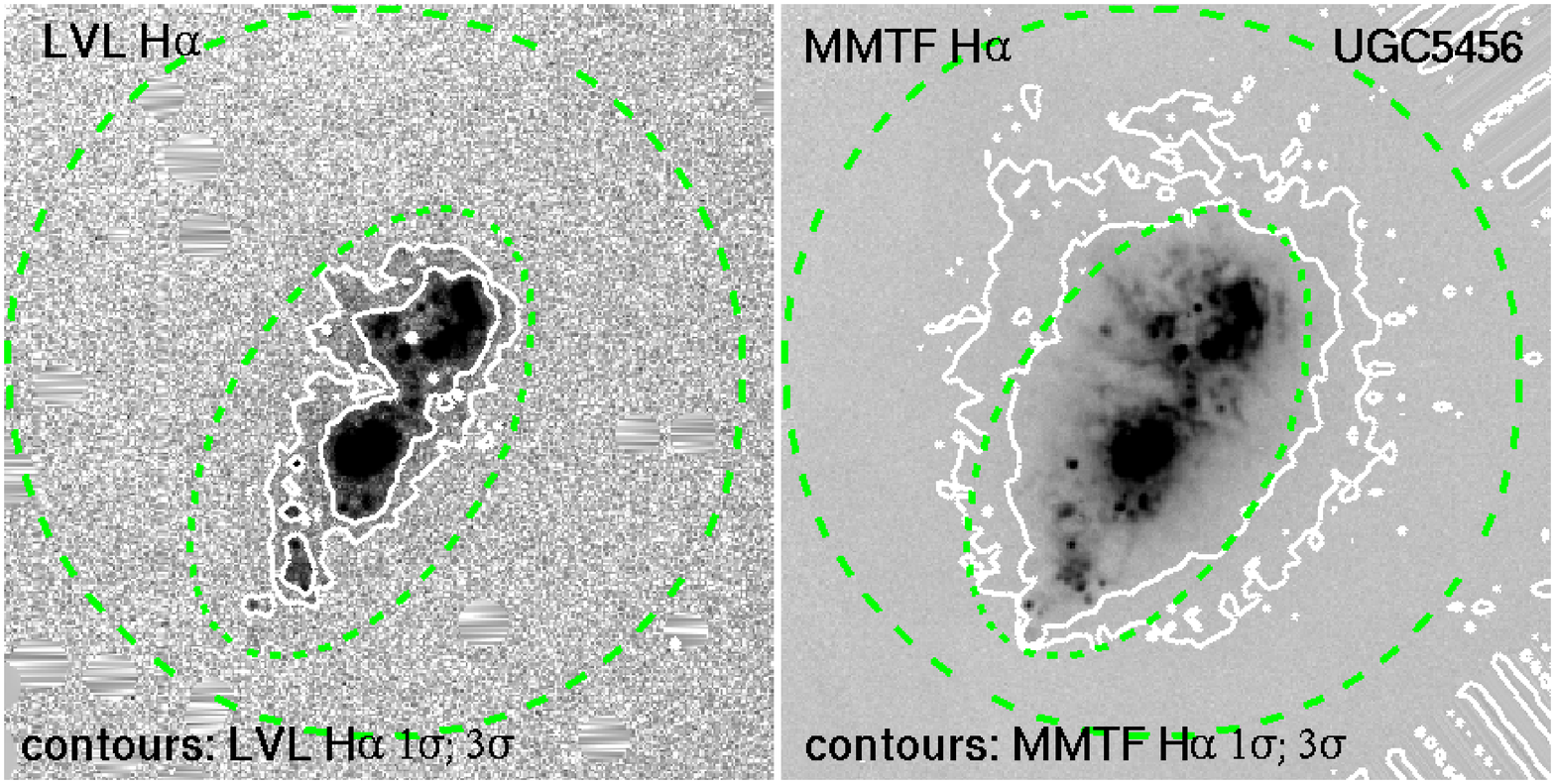}
\plotone{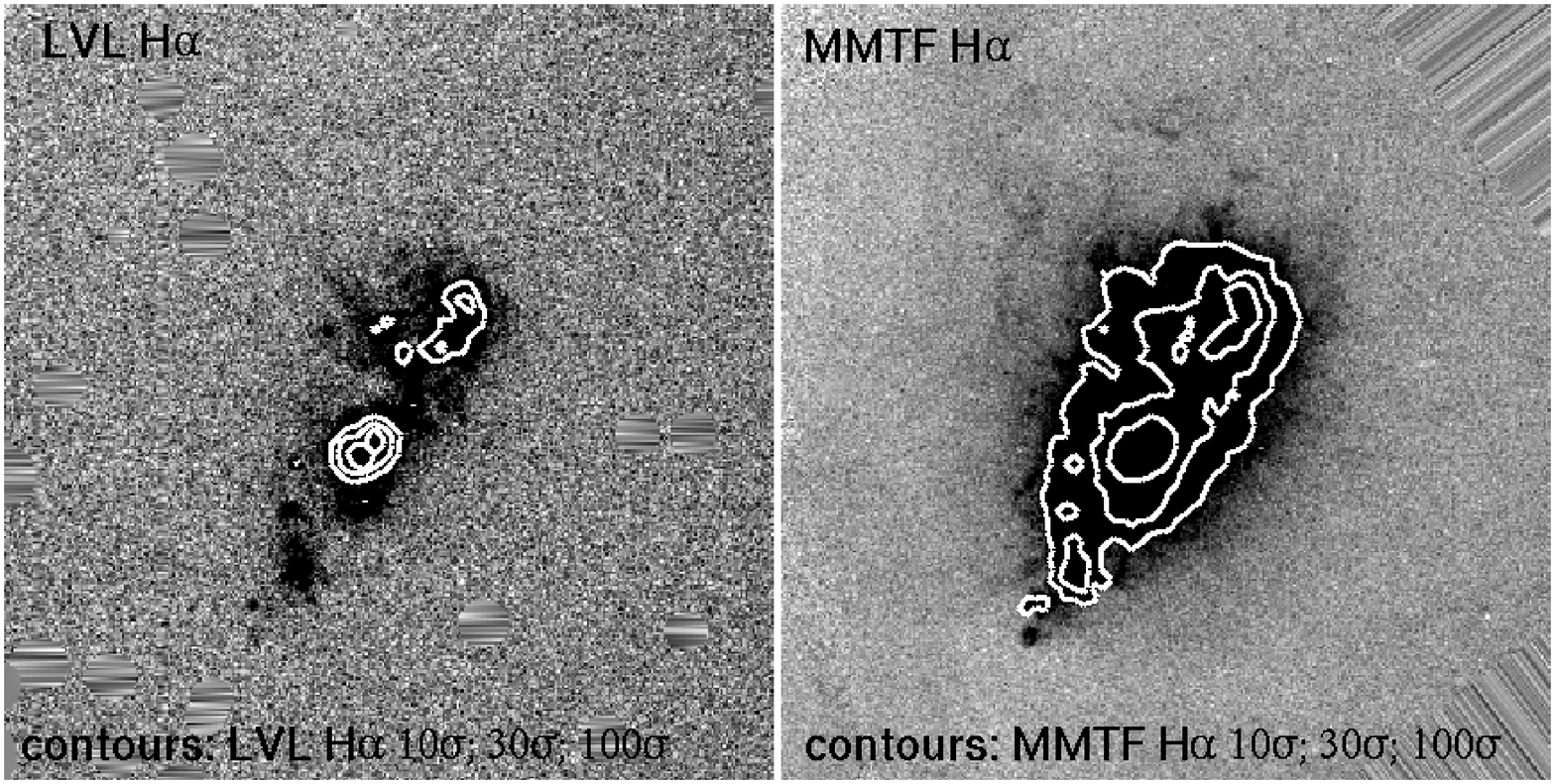}
\plotone{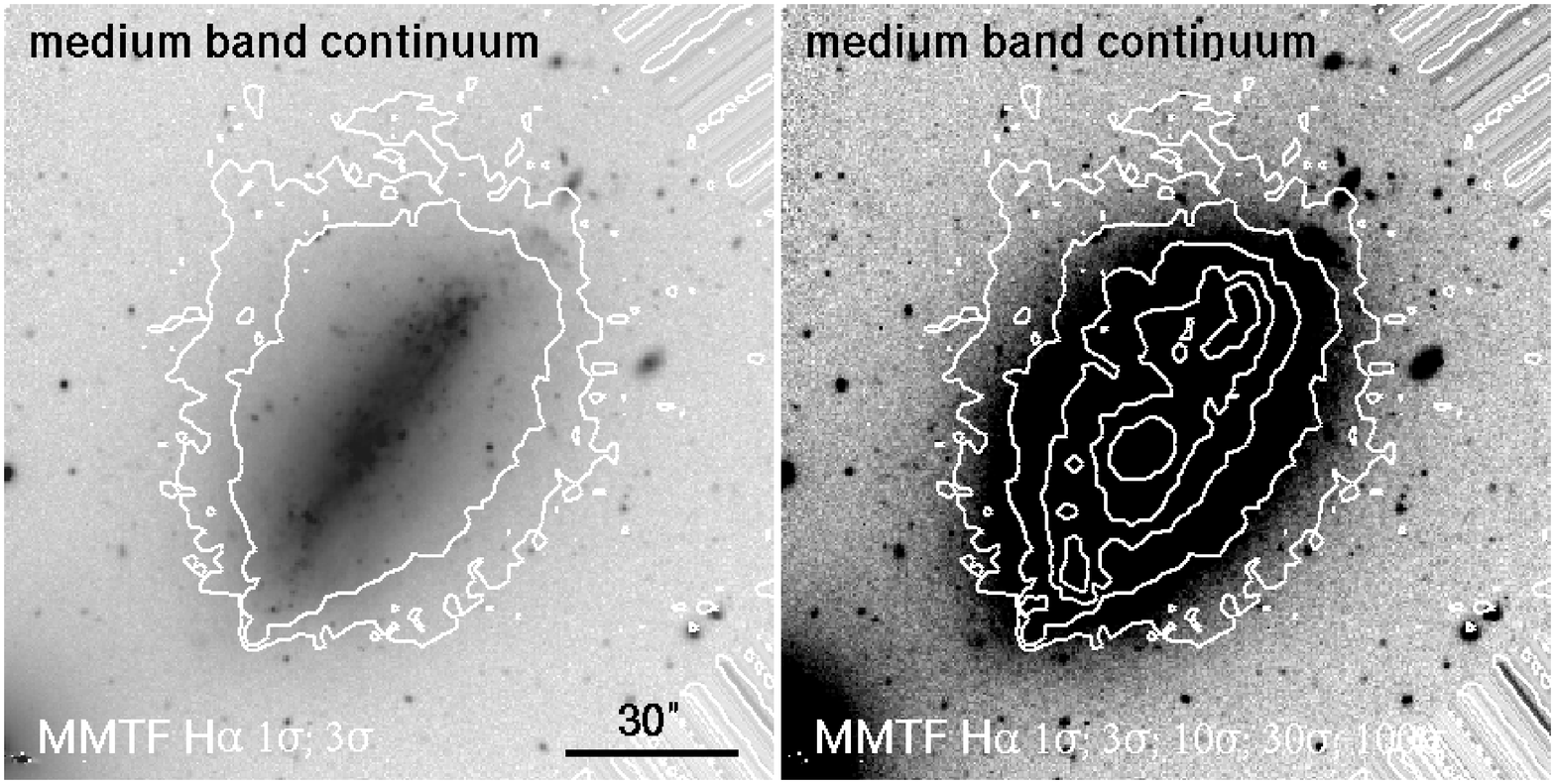}
\caption{Comparison of H$\alpha$ emission in UGC5456 detected through standard continuum subtracted narrowband imaging from the 11HUGS/LVL survey and through deeper observations with the MMTF as reported in this paper. The unit responses of the two images have been scaled to be consistent.  {\it Top row:} H$\alpha$ images are displayed on a logarithmic scale with a stretch which illustrates the morphology of HII regions and higher surface brightness features.  The contours shown are at flux levels 1$\sigma$ and 3$\sigma$ above the background, and are smoothed over four pixels.  1$\sigma$ corresponds to 4.4$\times 10^{-17}$ ergs s$^{-1}$ cm$^{-2}$ arcsec$^{-2}$ for the 11HUGS/LVL image, and to 4.8$\times 10^{-18}$ ergs s$^{-1}$ cm$^{-2}$ arcsec$^{-2}$ for the MMTF image.  Ellipses indicate the apertures through which the photometry reported in Table 1 is performed. {\it Middle row:} H$\alpha$ images are displayed on a linear scale and are stretched to emphasize emission from the diffuse ionized gas  Similarly, contours are shown, but now at 10$\sigma$,30$\sigma$, and 100$\sigma$ above the background.  {\it Bottom row:} H$\alpha$ contours from the MMTF imaging are over-plotted on the medium-band image used for continuum subtraction.  The same image is in both panels, but stretched to highlight the inner disk on the left (logarithmic scale) and the outer disk on the right (linear scale). Assuming a distance of 3.8 Mpc, the 30\arcsec\ bar in the lower left panel corresponds to 552 pc.}
\end {figure}

\begin{figure}
\epsscale{0.9}
\plotone{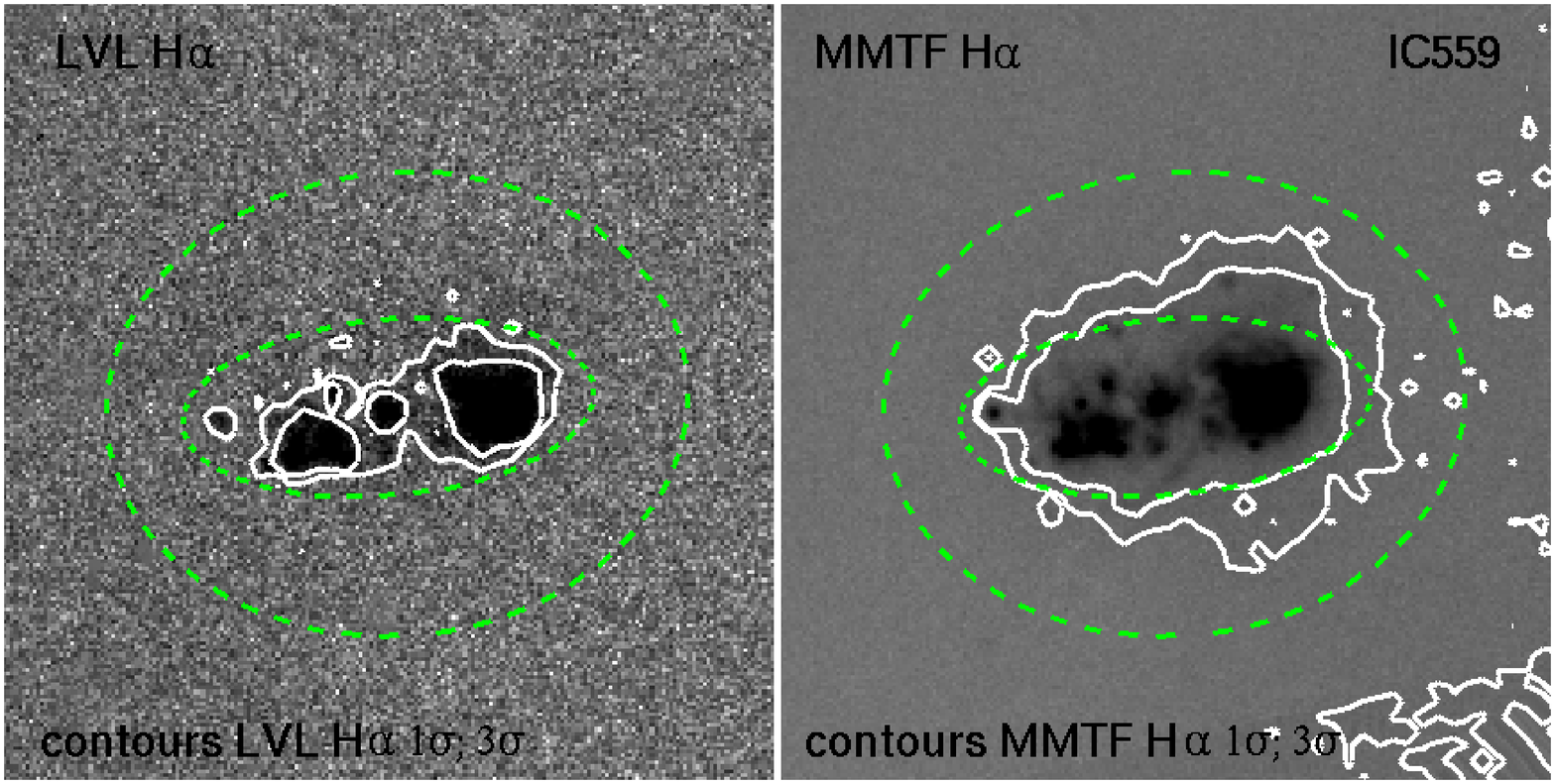}
\plotone{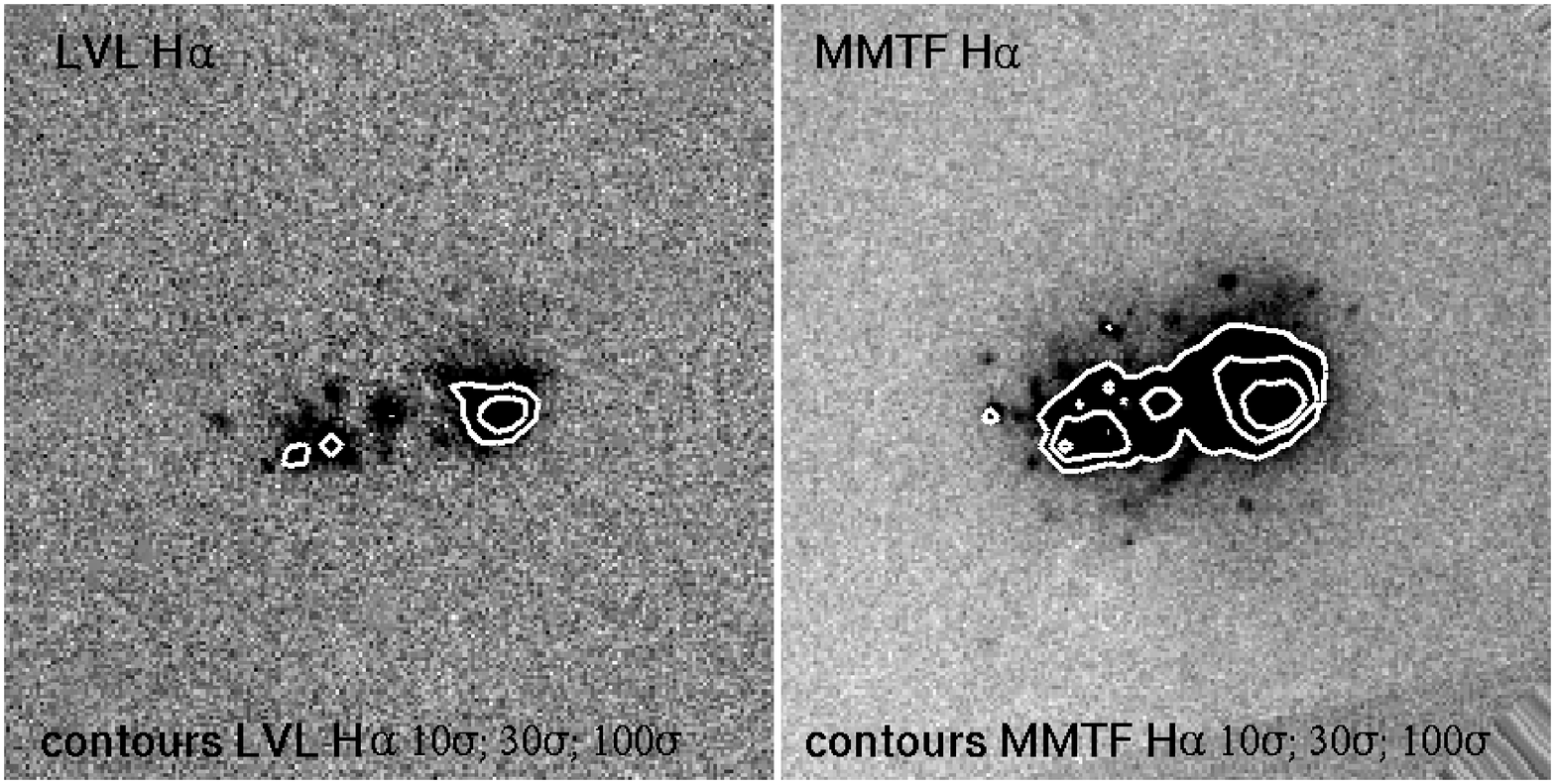}
\plotone{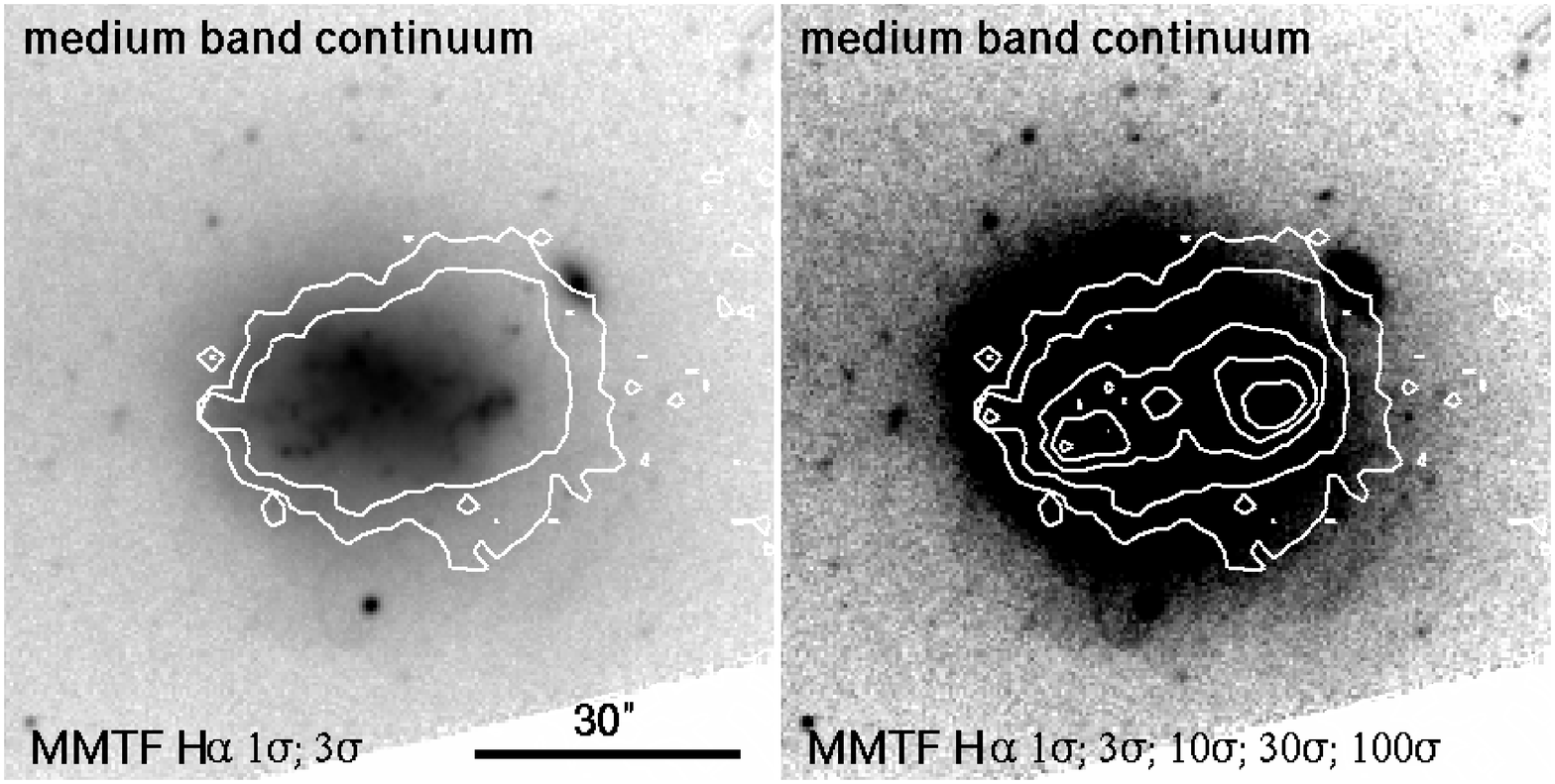}
\caption{Same as Figure 1, but for IC559.  1$\sigma$ corresponds to 3.2$\times 10^{-17}$ ergs s$^{-1}$ cm$^{-2}$ arcsec$^{-2}$ for the 11HUGS/LVL image, and to 4.1$\times 10^{-18}$ ergs s$^{-1}$ cm$^{-2}$ arcsec$^{-2}$ for the MMTF.  Assuming a distance of 4.9 Mpc, the 30\arcsec\ bar in the lower left panel corresponds to 710 pc.}
\end {figure}

\begin{figure}
\epsscale{0.9}
\plotone{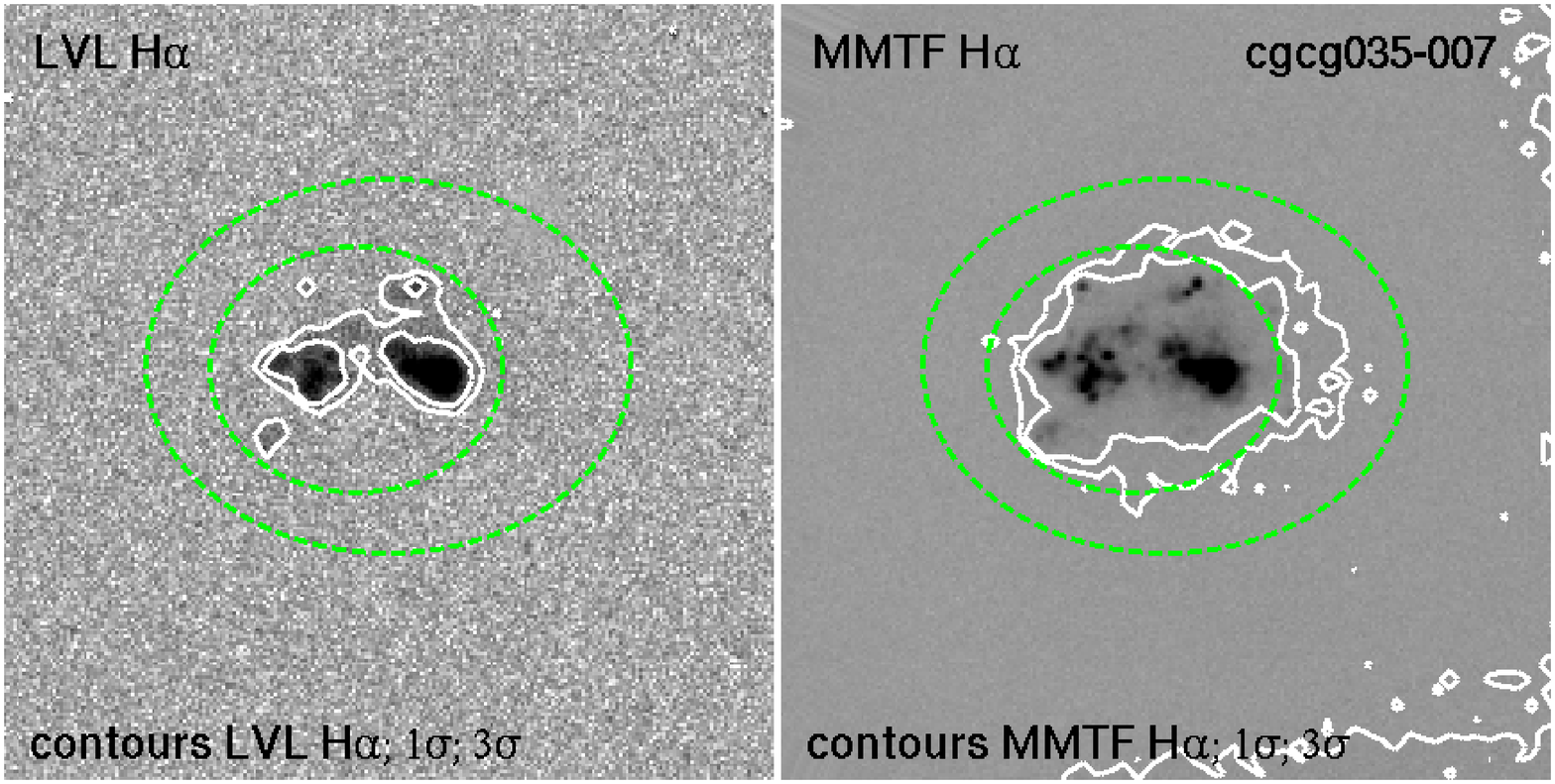}
\plotone{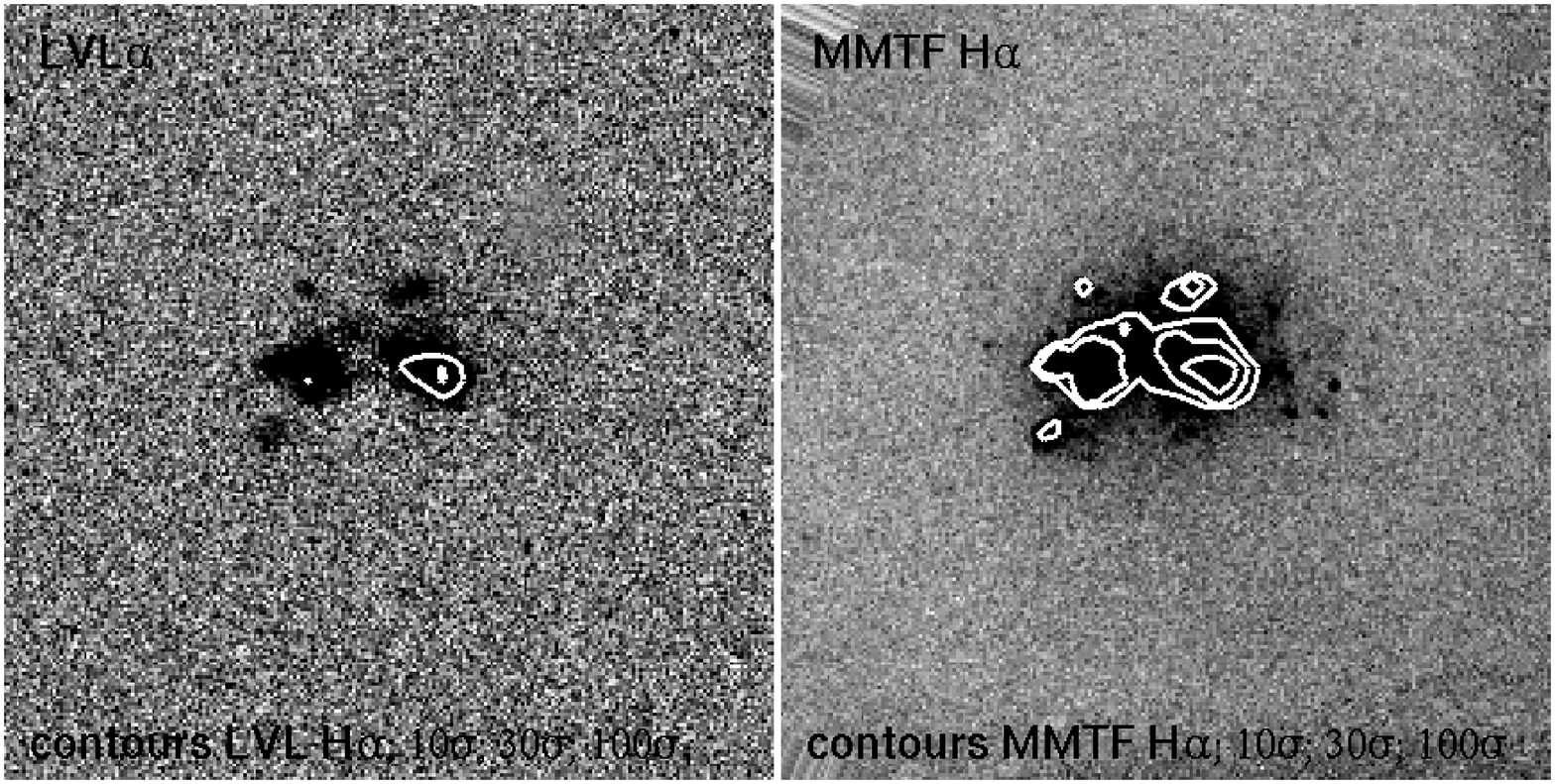}
\plotone{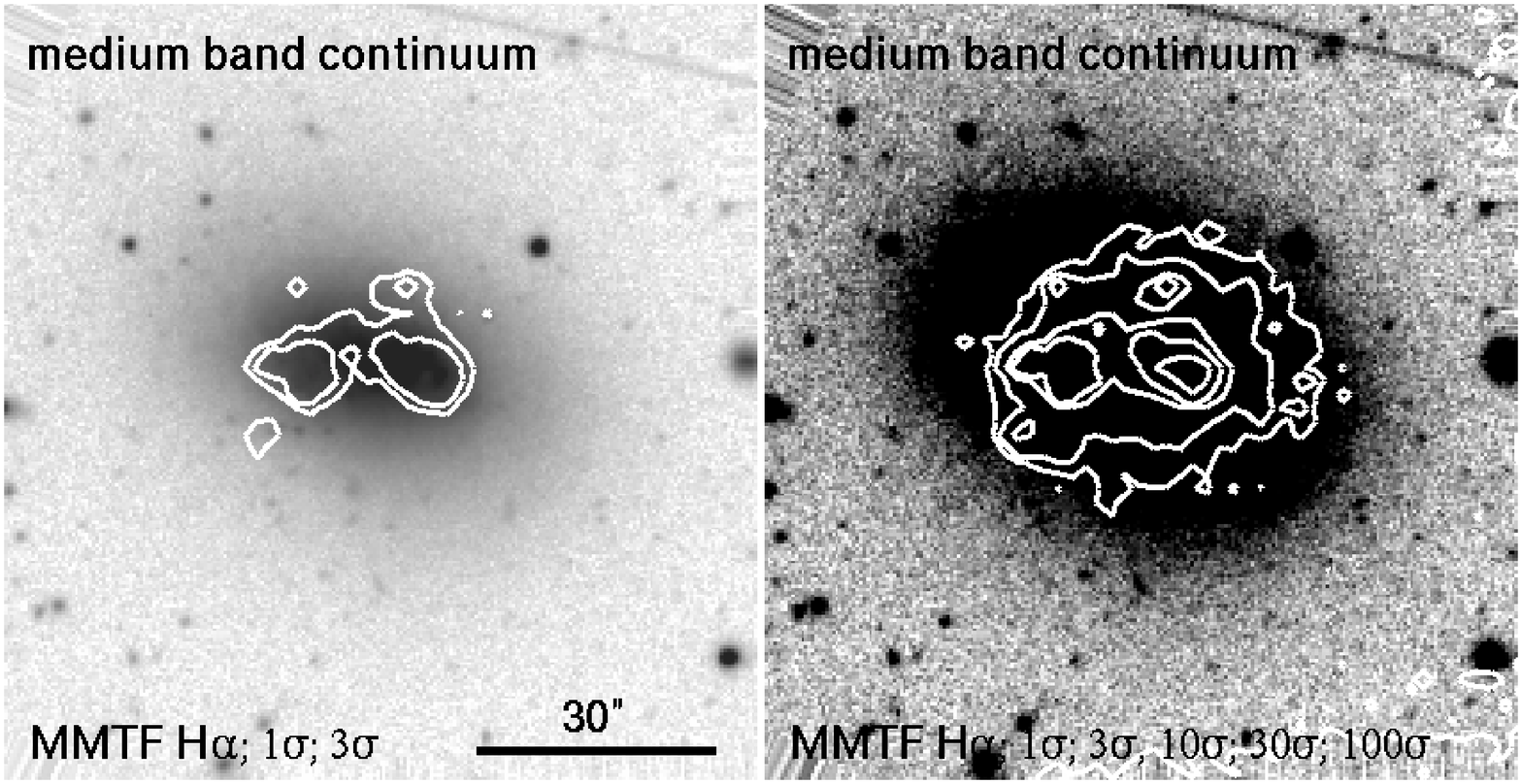}
\caption{Same as Figure 1 and 2, but for CGCG035-007.  1$\sigma$ corresponds to 3.4$\times 10^{-17}$ ergs s$^{-1}$ cm$^{-2}$ arcsec$^{-2}$ for the 11HUGS/LVL image, and to 4.4$\times 10^{-18}$ ergs s$^{-1}$ cm$^{-2}$ arcsec$^{-2}$ for the MMTF.  Assuming a distance of 5.2 Mpc, the 30\arcsec\ bar in the lower left panel corresponds to 760 pc.}
\end {figure}

\end{document}